\documentclass[epj]{svjour}
\usepackage{graphics}
\usepackage{epsfig}
\usepackage{amsmath}
\usepackage{amssymb}
\usepackage{hhline}
\usepackage{subfigure}

\newcommand{\pppKL}{\mbox{$pp \rightarrow pK^{+}\Lambda$}\ }
\newcommand{\pppKSo}{\mbox{$pp \rightarrow pK^{+}\Sigma^{0}$}\ }

\newcommand{\be}{\begin{equation}}
\newcommand{\ee}{\end{equation}}
\newcommand{\bea}{\begin{eqnarray}}
\newcommand{\eea}{\end{eqnarray}}
\newcommand{\nn}{\nonumber}

\begin{document}
\title{Energy dependence of the $\boldsymbol{\Lambda/\Sigma^0}$ production
cross section ratio in p--p interactions.}
\author{
P.~Kowina\inst{1,5}$^,$\thanks{{e-mail:} p.kowina@fz-juelich.de},
M.~Wolke\inst{1},    
H.-H.~Adam\inst{2},
A.~Budzanowski\inst{3},
R.~Czy$\dot{\mbox{z}}$ykiewicz\inst{4},
D.~Grzonka\inst{1},
J.~Haidenbauer\inst{1},
M.~Janusz\inst{4},
L.~Jarczyk\inst{4},
B.~Kamys\inst{4},
A.~Khoukaz\inst{2}, 
K.~Kilian\inst{1},
P.~Moskal\inst{1,4},
W.~Oelert\inst{1},
C.~Piskor-Ignatowicz\inst{4},
J.~Przerwa\inst{4},
C.~Quentmeier\inst{2},
T.~Ro$\dot{\mbox{z}}$ek\inst{5},
R.~Santo\inst{2},
G.~Schepers\inst{1},
T.~Sefzick\inst{1},
M.~Siemaszko\inst{5}, 
J.~Smyrski\inst{4}, 
S.~Steltenkamp\inst{2},
A.~Strza{\l}kowski\inst{4},
A.~T\"aschner\inst{2},
P.~Winter\inst{1},
P.~W{\"u}stner\inst{6},
W.~Zipper\inst{5}           
}                     % Do not remove
\institute{
Institut f{\"u}r Kernphysik, Forschungszentrum J\"{u}lich, D--52425 J\"ulich, Germany
\and
Institut f{\"u}r Kernphysik, Westf{\"a}lische Wilhelms--Universit{\"a}t,  D--48149 M{\"u}nster, Germany
\and
Institute of Nuclear Physics, PL--31 342 Cracow, Poland
\and
Institute of Physics, Jagellonian University, PL--30 059 Cracow, Poland
\and
Institute of Physics, University of Silesia, PL--40 007 Katowice, Poland
\and
Zentrallabor f{\"u}r Elektronik,  Forschungszentrum J\"{u}lich, D--52425 J\"ulich, Germany
}
\date{Received: \today / Revised version: date}
\abstract{
The production of the $\Lambda$ and $\Sigma^0$ hyperons has been measured via 
the $pp \rightarrow p K^+ \Lambda / \Sigma^0$ reaction at the internal 
COSY--11 facility in the excess energy range between 14 and 60 MeV. The 
transition of the $\Lambda/\Sigma^0$ cross section ratio from about 28 at 
$\mbox{Q} \le 13~MeV$ to the high energy level of about 2.5 is covered by 
the data showing a strong decrease of the ratio between 10 and 20 MeV excess 
energy.\\
Effects from the final state interactions in the  $p-\Sigma^0$ channel
seems to be much smaller compared to the $p-\Lambda$ one.
Estimates of the effective range parameters are given for the $N\Lambda$ and 
the $N\Sigma$ systems.
\PACS{
{13.75.-n}{Hadron-induced low- and intermediate-energy reactions and
                            scattering (energy(less-than-or-equal-to)10 GeV)} \and 
{13.75.Ev}{Hyperon-nucleon interactions} \and 
{13.85.Lg}{Total cross sections} \and 
{14.20.Jn}{Hyperons} \and 
{25.40.Ep}{Inelastic proton scattering} 
%{29.20.Lq}{Synchrotrons}  
%      {PACS-key}{discribing text of that key}   \and
%      {PACS-key}{discribing text of that key}
     } % end of PACS codes
} %end of abstract
\authorrunning{P.~Kowina et al.}
\titlerunning{Energy dependence of the $\Lambda/\Sigma^0$ production
cross section ratio}
\maketitle
\section{Introduction}
\label{introduction}

\begin{figure}[h]
\centerline{
\psfig{figure=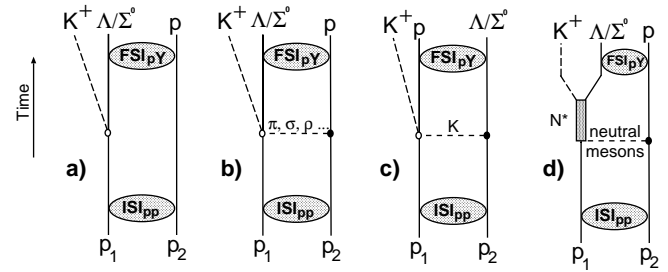,width=3.7cm,angle=-90}}
\vspace{0.2cm}
\caption[feyman]
{\label{feyman}
Examples of possible graphs for the $\Lambda$ or $\Sigma^0$ hyperon %($Y=\Lambda$ or $\Sigma^0$) 
production with inclusion of proton-proton initial (ISI) and proton-hyperon 
final (FSI) state interactions.
}
\end{figure}

In the kinematical threshold region the strangeness 
production is commonly described by both non-strange and 
strange meson exchange with or without explicit inclusion of
an intermediate resonance as depicted in the four graphs of Fig.~\ref{feyman}. 
While the exchange of the lightest mesons, namely $\pi$ and $K^+$, is expected
to be dominant in the $\Lambda$ and $\Sigma^0$ 
production~\cite{fael97,li95,li97,sib198,lag91,lag01,gas00}
there could be also a contribution arising from the exchange of heavier 
non-strange or strange mesons~\cite{tsu97,tsu99,sib00b,shy01}. In addition,
proton-hyperon final state interactions (FSI) play an important role when 
comparing the $pK^+\Lambda$ and $pK^+\Sigma^0$ reaction channels.
\vspace{0.1cm}

Since the quark structures of the two neutral $\Lambda$ and $\Sigma^0$ hyperons
are analogous to each other one can expect similar production 
mechanisms. 
In such a case the cross section ratio ${\cal{R}}_{\Lambda/\Sigma^0} 
\equiv \frac{\sigma(pp \to pK^+\Lambda)}{\sigma(pp \to pK^+\Sigma^0)}$ 
should be mainly determined by the isospin relation which leads  
to ${\cal{R}}_{\Lambda/\Sigma^0} \sim 3$, in good agreement with the value 
${\cal{R}}_{\Lambda/\Sigma^0} \approx 2.5$ observed in proton-proton 
scattering experiments at excess energies 
$\mbox{Q} \ge 300\,\mbox{MeV}$~\cite{bal88}. It is interesting to note that
a comparable cross section ratio was determined in antiproton-proton 
annihilation experiments leading to $\bar\Lambda \mbox{-} \Lambda$,  
$\bar{\Sigma^0} \mbox{-} \Lambda + c.c.$ and 
$\bar\Sigma^{\pm} \mbox{-} \Sigma^{\mp}$ as performed by the PS185 
collaboration at LEAR~\cite{barn97,barn97b}.

	In contrary to these results, close to threshold data~\cite{sew99}
(i.e.\ $\mbox{Q} \le 13 \,\mbox{MeV}$) taken at the COSY--11 
facility~\cite{bra96,pawel:nim} revealed as the most remarkable characteristic 
a ratio of ${\cal{R}}_{\Lambda/\Sigma^0} \thickapprox 28$.
% where the total cross sections were compared each at the same excess energy.
%
%        However, the most remarkable feature of the 
%$pp~\rightarrow p K^+ \Lambda\,(\Sigma^0)$ data~\cite{sew99} measured 
%at the COSY--11 facility~\cite{bra96,pawel:nim} in the very close to 
%threshold range i.e.\ for excess energies $\mbox{Q} \le$13~MeV
%was that at the same excess energy the total cross section for the 
%$\Sigma^0$ production is by a factor of $28^{\,+6}_{\,-9}$ smaller than for 
%the $\Lambda$ particle. 

       At first a dominant $K$-exchange mechanism was discussed~\cite{sew99} 
as a possible explanation of the large $\Lambda/\Sigma^0$ cross section ratio,
which can be 
determined by the ratio of the squared  coupling constants 
$g_{\Lambda N K}^2 / g_{\Sigma N K}^2$, assuming that all other components
in the $\Lambda$  and $\Sigma^0$ production diagrams are the same and when
neglecting any final state interactions. 
%However, since the ratio of the 
%coupling constants is not well known and varies between 0.08 
%and 27~\cite{li95,li97,lag91,lag01,sib198,del89,mae89,ant86,mar81,dal82,sie94,hol89}, 
%a significant difference in those model predictions is implicated.
It is particularly interesting to note that the ratio 
of the coupling constants within the SU(6) representation~\cite{swa63,dov84} 
leads to ${\cal{R}}_{\Lambda/\Sigma^0}$~=~27, which exactly reproduces the 
experimental result 
very close to threshold 
but fails to explain the cross section ratio observed
at high excess energies. In reality it is expected that also  $\pi$-exchange
contributes significantly in the production process. For a review of that
issue the reader is refered to~\cite{revart}.
      
        The first COSY--11 data on the ratio encouraged several 
groups to explain the experimental results. 
%and different theoretical investigations started. 
For example calculations within a meson exchange model~\cite{gas00} taking 
into account pion and kaon exchange (graphs b)~and~c) in Fig.~\ref{feyman}) 
and their interference reproduce the measured excitation functions of
$\Lambda$ and $\Sigma^0$ production: 
        while the $\Lambda$ production channel appears to be dominated by the 
$K$-exchange mechanism, both $\pi$ and $K$ exchange contribute with
equal strength to the $\Sigma^0$ production.
Therefore a~$\pi$-$K$ interference becomes significant only in the $\Sigma^0$
case. Indeed the authors of Ref.~\cite{gas00} argue that the data require 
--- independent of the hyperon-nucleon potential used for the description of 
the low-energy final state scattering process --- a destructive interference 
between the $\pi$ and $K$ exchange contributions. 

      Other recent models~\cite{lag91,lag01,sib00b,shy01} describe the 
ratio {$\cal{R}$} within a factor of two by including heavier exchange mesons 
(graph b) Fig.~\ref{feyman}) or nucleon resonances 
(graph d) Fig.~\ref{feyman}).

        More data were needed to differentiate between the various 
descriptions and to develop a suitable model for the hyperon production.
Therefore, additional cross sections were measured at the COSY--11 facility in
order to determine the energy dependence of the $\Lambda/\Sigma^0$ cross 
section ratio in the transition region from $Q \approx 14\,\mbox{MeV}$ to 
$Q \approx 60\,\mbox{MeV}$ where the production ratio changes  most 
drastically.

\section{Experiment\label{experiment}}

\begin{figure}[h]
%\vspace{-.8cm}
\centerline{
\psfig{figure=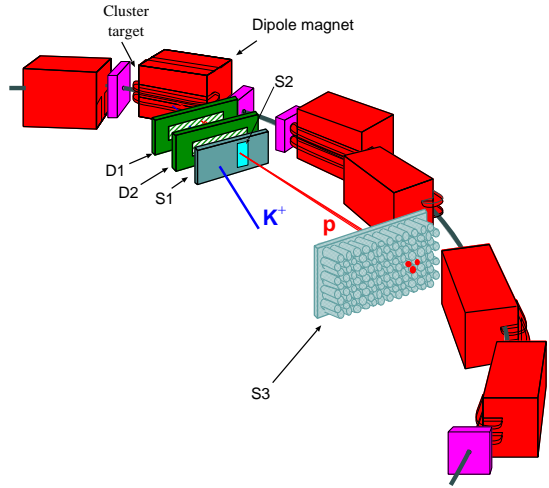,width=0.4\textwidth,angle=0}}
\vspace{0.1cm}
\caption[cosy11]
{\label{cosy11}
COSY--11 detection facility.
}
\end{figure}

The present hyperon (Y) production experiments were performed 
at the {\bf{Co}}oler {\bf{Sy}}n\-chrotron COSY-J\"ulich~\cite{maie97}
using the COSY--11 detection facility~\cite{bra96,pawel:nim} shown in 
Fig.~\ref{cosy11}.

	One of the regular COSY dipole magnets serves as a magnetic 
spectrometer with a H$_{2}$ cluster beam target~\cite{domb97} installed in 
front of it.
%	Due to the relatively small target density of $\sim$\ 5$\cdot10^{~13}$ 
%atoms/cm$^3$  energy losses and 
%secondary reactions in the target are negligible. 
Due to the advantage of internal beam experiments to use targets with 
comparatively low areal densities (e.g. $\sim$\ 5$\cdot10^{~13}$ atoms/cm$^3$),
energy losses and secondary reactions in the target are negligible. The 
interaction between a proton of the beam with a proton of the cluster target 
may lead to the production of neutral hyperons ($\Sigma^0$, $\Lambda$) via the 
reactions $pp \to pK^+\Lambda(\Sigma^0)$.
%$\pppKL or \pppKSo$. 

	 The selection of $pK^+\Lambda(\Sigma^0)$ events is done
by the detection of both positively charged particles in the exit channel 
(i.e.\ proton and $K^{+}$) while the unobserved neutral particle 
is identified via the missing mass method.

Positively charged ejectiles from these reactions have 
smaller momenta than the protons in the beam and therefore, they are directed
from the circulating beam by the magnetic field of the dipole towards the 
inner part of the COSY ring.

Leaving the vacuum chamber through a very thin exit foil 
(300 $\mu$m carbon fibre/30 $\mu$m Al)~\cite{bra96}, the 
charged reaction products are registered in a set of two drift chambers D1 
and D2 for the track reconstruction. With the well known 
magnetic field, their momenta can be determined by tracking back to the
interaction point. Furthermore, a time-of-flight measurement between 
the S1(S2) start and the S3 stop scintillator hodoscopes allows to determine
their velocity. Therefore, the four-momentum vectors for all positively 
charged particles  can be established and together with the known initial 
kinematics, the four-momentum of the unobserved neutral hyperon is uniquely 
determined.

	To avoid systematical uncertainties as much as possible, COSY was 
operated in the ''supercycle mode'' i.e.\ the beam momenta were changed 
between the cycles, such that for example 10 cycles with a beam momentum 
corresponding to the excess energy $Q=20$~MeV above the $\Sigma^0$ threshold 
were followed by one cycle with the same $Q$ above the $\Lambda$ production 
threshold. The ratio of the number of the cycles was chosen inverse to the 
ratio of the cross sections for the $\Lambda$ and $\Sigma^0$ production based 
on the expectations from previous experiments. Thus, both cross sections were 
measured under the same conditions and possible changes in the detection 
system did not influence the data taking procedure, especially for the 
determination of the cross section ratio.

\begin{figure}[h]
\vspace{-0.9cm}
\centerline{
\psfig{figure=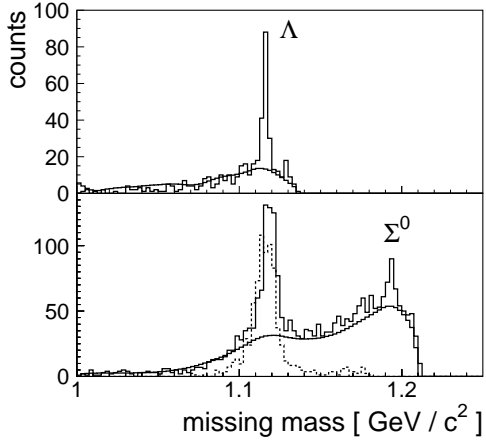,width=0.4\textwidth,angle=0}}
\vspace{0.1cm}
\caption[supercycle]
{\label{supercycle}
Spectra of the missing mass for the \pppKL (upper part) and \pppKSo (lower 
part) reactions measured at the same excess energy $Q=20~\mbox{MeV}$ above the 
$\Lambda$ and $\Sigma^0$ threshold, respectively.
}
\vspace{-0.3cm}
\end{figure}

        In Fig.~\ref{supercycle} typical missing mass spectra for the threshold 
production of $\Lambda$  and $\Sigma^0$ hyperons are given, both measured at 
the same excess energy of $Q$~=~20~MeV. It is obvious that the peak to 
background ratio is much larger for the $\Lambda$ than for the $\Sigma^0$ 
production. The solid lines are background distributions as derived from the 
experimental data themselves by selecting side bands of the two dimensional 
invariant vs. missing mass representation below and above the kaon range as 
described in a previous publication~\cite{sew99}.  
The small excess of counting
rate in the lower part of the figure at smaller missing masses than the 
$\Sigma^0$ peak is be due to $pp \to pK^+ \Lambda \gamma$ and 
$pp \to pK^+ \Lambda$ reactions with subsequent $\Lambda$ decay and 
missidentification of the primary and secondary protons. The dashed 
line in the lower panel of Fig.~\ref{supercycle} shows the distribution for  
the $pp \to pK^+ \Lambda$ reaction obtained in Monte-Carlo simulations.
The background shoulder beyond the $\Lambda$ peak in the figure for the 
$\Sigma^0$ hyperon production is predominantly due to the decreasing missing 
mass resolution with increasing $Q$ value~\cite{jurek:annrep}.

\vspace{-0.15cm}
\section{Results\label{results}}
\vspace{-0.15cm}
\subsection{Total cross section}

      The total cross sections extracted for all measured beam momenta are 
listed in Tables~\ref{cross_lambda} and \ref{cross_sigma} where only the 
statistical errors are given. In addition, the following systematic 
uncertainties  have to be taken into account: 
for the background subtraction 15~\% for $\Sigma^0$ and 7~\% 
for $\Lambda$ as well as 3.5~\% due to the
luminosity determination which is performed by a simultaneous measurement
of proton--proton elastic scattering~\cite{pawel:nim}.
 A further systematical error of below 5~\% is introduced by the detection 
efficiency which is determined via GEANT~\cite{geant} simulations assuming 
an S-wave production process with inclusion of the proton-$Y$ FSI only in 
the case of the \pppKL reaction. The detection efficiencies for different
access energies are shown in Tables~\ref{cross_lambda} and~\ref{cross_sigma}.
To deduce the influence of possible higher partial wave contributions a P-wave
in the $p$-$Y$ system  was introduced in the MC simulations. Compared to S-wave
pure P-wave production would result in an increase of the total acceptance of 
about a factor two for the highest $Q = 60 \mbox{MeV}$. This moderate 
difference can be explained by the special configuration 
of the detection system with a nearly full acceptance in the horizontal plane
resulting in the full coverage of the $\Theta$ angular range for a thin slice 
in the vertical plane.

	In reality P-wave contribution is expected to be below 20\% at  
a $Q$-value of $60\: \mbox{MeV}$~\cite{tof:blig98} and much smaller at 
lower $Q$ values. Therefore, an additional systematical error in acceptance 
calculation due to the higher partial wave in the $p$-$Y$ system would be
below 20\%.

%\vspace{-0.2cm}
\begin{table}[h]
\vspace{-.3cm}
 \begin{center}
\caption{
Values of the cross sections for the $\Lambda$ production in proton-proton 
collisions for six excess energies above the production threshold.
}
\label{cross_lambda}
   %\hspace{-1.0cm}
   \begin{tabular}{|c|c|c|}
   \hline
\multicolumn{3}{|c|}{\pppKL}\\
\hline        
 Excess   energy    &  Total cross section &Detection efficiency \\
 $ Q$ \mbox{} [MeV] & $\sigma$ \mbox{} [nb]& {[\%]}\\
\hline
 13.9 $\pm$ 0.5      &  630 $\pm$ 79 & 0.65  $\pm$ 0.02 \\
 15.9 $\pm$ 0.7      &  727 $\pm$ 57 & 0.56  $\pm$ 0.015\\
 20.2 $\pm$ 0.7      & 1011 $\pm$ 99 & 0.39  $\pm$ 0.01 \\
 30.1 $\pm$ 0.5      & 1366 $\pm$ 247& 0.25  $\pm$ 0.01 \\
 39.7 $\pm$ 1.1      & 2118 $\pm$ 266& 0.145 $\pm$ 0.005\\
 59.3 $\pm$ 0.5      & 3838 $\pm$ 624& 0.075 $\pm$ 0.005\\
\hline
\end{tabular}
\end{center}
\vspace{-1.1cm}
\end{table}

%\vspace{-0.3cm}
\begin{table}[h]
\caption{
Values of the cross sections for the $\Sigma^0$ production in proton-proton 
collisions for six excess energies above the production threshold.
}
\label{cross_sigma}
 \begin{center}
   \begin{tabular}{|c|c|c|}
   \hline
\multicolumn{3}{|c|}{\pppKSo}\\
\hline        
 Excess   energy    &  Total cross section &Detection efficiency \\
 $ Q$ \mbox{} [MeV] & $\sigma$ \mbox{} [nb]& {[\%]}\\
\hline
 13.8 $\pm$ 0.5      & 34.9 $\pm$ 6.5& 0.76  $\pm$ 0.02  \\
 15.9 $\pm$ 0.9      & 46.8 $\pm$ 6.4& 0.66  $\pm$ 0.015 \\
 20.3 $\pm$ 0.7      &   78 $\pm$ 14 & 0.47  $\pm$ 0.01  \\
 29.9 $\pm$ 0.5      &  125 $\pm$ 32 & 0.28  $\pm$ 0.01  \\
 39.7 $\pm$ 1.3      &  196 $\pm$ 33 & 0.167 $\pm$ 0.005 \\
 59.1 $\pm$ 0.5      &  482 $\pm$ 144& 0.078 $\pm$ 0.005 \\
\hline
\end{tabular}
\end{center}
\vspace{-0.4cm}
\end{table}

  Previous COSY--11 measurements of the  excitation function
for the \pppKL and \pppKSo reactions at excess energies up to 
$Q$~=~13~MeV above the production threshold are summarized in~\cite{sew99},
where it was pointed out that the energy dependence of the cross section is 
much better described by a 3-body phase space behaviour modified by the 
proton-hyperon ($p-Y$) final state interaction~\cite{fael97} than by a 
calculation taking into account pure phase space only.
	However, it should be emphasized, that in~\cite{sew99} these fits were 
naturally limited to the excess energy range of the available data, i.e. 
$Q$~$\le$~13~MeV.

\begin{figure}[h]
\vspace{-0.8cm}
\centerline{
\psfig{figure=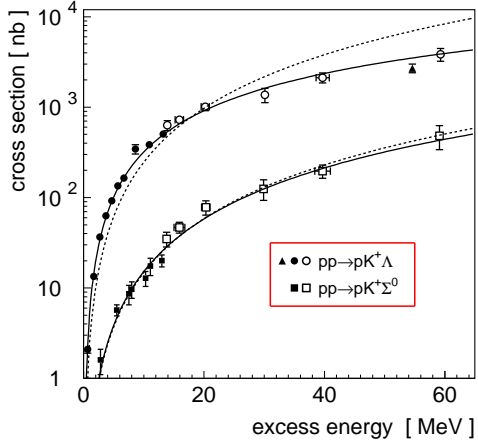,width=0.4\textwidth,angle=0}}
\vspace{0.0cm}
\caption[cross_section_plot]
{\label{cross_section_plot}
Total cross sections for the \pppKL and \pppKSo production. Open symbols 
present results of this work. Full symbols were determined from previous 
measurements of the COSY--11~\cite{sew99,bal98a} (circles and squares) and 
COSY-TOF collaboration~\cite{tof:blig98} (triangle). The plotted error bars
include only statistical errors.
}
\vspace{-0.2cm}
\end{figure}

        In Fig.~\ref{cross_section_plot} the first COSY--11 
data~\cite{sew99,bal98a} and one data point obtained by the COSY--TOF 
collaboration~\cite{tof:blig98} are now extended by the new results up to 
$Q$~=~60~MeV.

        The dashed lines in Fig.~\ref{cross_section_plot} represent 
$\chi^2$ fits to the full set of data points in the excess energy range 
%$14~\mbox{MeV} \leq 
$Q \leq 60~\mbox{MeV}$ with a pure
phase space behaviour~\cite{byckling}:
\vspace{-0.25cm}
\begin{equation}\label{PhS_eqation}
\sigma = K \cdot Q^2,
\end{equation}

\vspace{-0.25cm}
\hspace{-.36cm}with the normalisation constant $K$ resulting in:\vspace{0.1cm}

\hspace{2cm}$K(\Lambda)$ =  (3.08~$\pm$~0.06)  nb/MeV$^2$ \vspace{0.1cm}

\hspace{2cm}$K(\Sigma^0)$ = (0.150~$\pm$~0.008) nb/MeV$^2$.\vspace{0.2cm}

	The solid lines in Fig.~\ref{cross_section_plot} are $\chi^2$ fits  
by the F\"aldt and Wilkin parametrisation ~\cite{fael97}
with inclusion of the energy dependent flux factor $F$ which accounts for the
different kinematics~\cite{revart}
\vspace{-0.2cm}        
\bea
\nn
\sigma \;=\; 
const\cdot \frac{V_{ps}}{\mbox{F}} \cdot 
  \frac{1}
    {\left(1\;+\;\sqrt{1\,+\,\frac{\mbox{\scriptsize Q}}{\epsilon^\prime}}\right)^2} 
\eea
$~~$\\[-0.6cm]
\be\label{faldtwilkinflux}
=\;C^{\prime} \cdot 
  \frac{\mbox{Q}^{2}}{ \sqrt{\lambda(\mbox{s},\mbox{m}_p^2,\mbox{m}_p^2)}} 
  \cdot 
  \frac{1}
    {\left(1\;+\;\sqrt{1\,+\,\frac{\mbox{\scriptsize Q}}{\epsilon^\prime}}\right)^2}\;.
\end{equation}
The phase space volume $V_{ps}$ and the flux factor F are 
given by~\cite{byckling}

\begin{equation}\label{Vps_nonrelativistic}
V_{ps} = \frac{\pi^{3}}{2} 
  \frac{\sqrt{\mbox{m}_{p}\,\mbox{m}_{K^+}\,\mbox{m}_{Y}}}
       {(\mbox{m}_{p} + \mbox{m}_{K^+} + \mbox{m}_{Y})^\frac{3}{2}} \; \mbox{Q}^2,
\end{equation}
\vspace{2ex}
$~~$\\[-1.1cm]
\begin{equation}\label{fluxfactor}
\mbox{F} = 
  2\,(2\pi)^{3n-4}\;\sqrt{\lambda(\mbox{s},\mbox{m}_p^2,\mbox{m}_p^2)}.
\end{equation}
The triangle function $\lambda$ is defined, e.g., in~\cite{byckling}.
 The parameter $\epsilon^{\prime}$,\ which is related to the strength of the 
$p-Y$ final state interaction, and the normalisation constant $C^{\prime}$ are
determined by the $\chi^2$ fits for each reaction separately resulting in:
\vspace{0.2cm}

\hspace{2cm}$C^{\prime}( \Lambda)$    =        (98.2~$\pm$~3.7) nb/MeV$^2$ \vspace{0.15cm}

\hspace{2cm}$\epsilon^{\prime} (\Lambda)$  = ($5.51^{\,+0.58}_{\,-0.52}$) MeV \vspace{0.15cm}

\hspace{2cm}$C^{\prime}(\Sigma^0)$   =        (2.97~$\pm$~0.27) nb/MeV$^2$  \vspace{0.15cm}

\hspace{2cm}$\epsilon^{\prime} (\Sigma^0)$ = ($133^{\,+108}_{\,-44}$) MeV. \vspace{0.2cm}

	In case of the $\Lambda$ production the $\chi^2$-value for the fit 
is $\chi^2$~=~1.8, which demonstrates a much better description of 
the data than by pure phase space representation with $\chi^2$~=~ 27.

        Contrary, in case of $\Sigma^0$ production there is almost no 
difference between the fit results assuming phase space either with or 
without FSI, resulting in $\chi^2$~=~1.16 and $\chi^2$~=~1.11, respectively. 
This might indicate a rather weak $p-\Sigma^0$ FSI\footnote{
                        With $\epsilon^{\prime}\to\infty$ 
			in Eq.~\eqref{faldtwilkinflux} 
                        one obtains the pure phase space distribution given 
                        by Eq.~\eqref{PhS_eqation}.} although this could
be also feigned by either higher partial wave contributions
or an energy dependence of the elementary amplitude~\cite{gas01}.

        Having extracted $\epsilon^{\prime}$ from the fit to the data it is 
possible to express the results in terms of the averaged effective range 
parameters, namely the scattering length~$\hat a$ and the effective 
range~$\hat r$. Assuming only S-wave production, the $p-\Lambda (\Sigma^0)$
systems can be described using the Bargmann potentials~\cite{newt82}, where 
$\hat a$ and $\hat r$ are expressed as:
\begin{equation}\label{scat_par_colin}
  \hat a = \frac{\alpha + \beta}{\alpha \beta}\:, \quad \quad
  \hat  r = \frac{2}{\alpha + \beta}\:.
\end{equation}

While $\beta$ is a shape parameter, the value of $\alpha$ is
calculated via $\epsilon^{\prime}~=~\alpha^2
/ 2 \mu$ where $\mu$ is the reduced mass of the 
($p-Y$) system~\cite{newt82}. The sign of $\alpha$ is ambiguous but the 
negative value is chosen since (at least for $p-\Lambda$) an attractive 
interaction is expected~\cite{hol89,Rij:1998}.%\cite{haidpriv}.

\begin{figure}[h]
\vspace{-0.8cm}
\centerline{
\psfig{figure=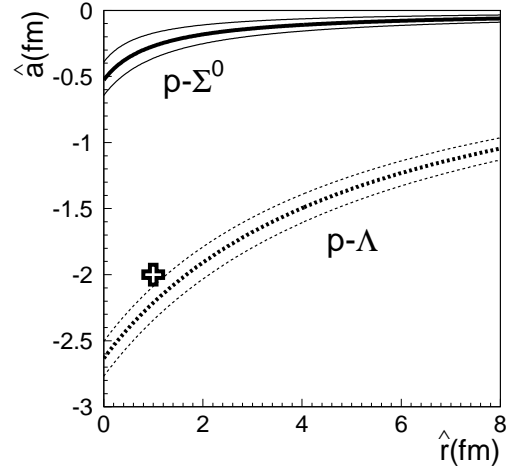,width=0.43\textwidth,angle=0}}
\vspace{0.0cm}
\caption[a_od_r_plot]
{\label{a_od_r_plot}
Correlation between the $p-\Sigma^0$  (solid lines) and $p-\Lambda$ 
(dashed lines) effective range parameters obtained as described in the text 
from  $\epsilon (\Sigma^0)$  and $\epsilon (\Lambda)$, respectively.
The cross symbol represents the averaged value of the $p-\Lambda$ effective 
range parameters extracted from a FSI approach in threshold $\Lambda$ 
production~\cite{bal98b}. 
%The full symbols are calculations for the $p$-$\Sigma^0$ system for different
%versions of the J\"ulich meson exchange model~\cite{haidpriv}.\\
}
\vspace{-0.3cm}
\end{figure}

As can be seen from Eq.~\eqref{scat_par_colin} the parameters $\hat a$ and 
$\hat r$ are interdependent and only a deduction of the correlations between 
them is possible. The extracted correlations for the $p-\Sigma^0$ and 
$p-\Lambda$ systems are presented in Fig.~\ref{a_od_r_plot} by solid and 
dashed lines, respectively. The error ranges (thinner lines) result from the
errors in $\epsilon^{\prime}$.

%The correlation of the effective range parameters 
The results obtained for the $p-\Lambda$ system are consistent with the value 
of the spin averaged parameters determined experimentally~\cite{bal98b} and 
represented in Fig.~\ref{a_od_r_plot} by the cross. In the $p-\Sigma^0$ 
system, the scattering length seems to be much smaller.
%The full symbols mark scattering parameters resulting from calculations
%within different versions of the J\"ulich meson exchange model \cite{haidpriv}
%in agreement with the correlation determined from the experiment.
However, one should be cautious with the interpretation of
the result for $p-\Sigma^0$. Because of possible transitions
in the final state (like $n\Sigma^+ \to p\Sigma^0$) the
use of the F\"aldt and Wilkin parametrization allows only to
obtain a qualitative estimation of the FSI effects induced by the 
$N-\Sigma$ interaction but not directly on the $p-\Sigma^0$ channel.
%~\cite{haidpriv}.

\subsection{Energy dependence of the  $\Lambda/\Sigma^0$ cross 
section ratio}  
\label{cross_section_ratio}
 
	The cross section ratios for the excess
energy range \\$14~\mbox{MeV} \leq Q \leq 60~\mbox{MeV}$ are listed in 
Table~\ref{cross_ratio}$\,$(and graphically presented in 
Fig.~\ref{cross_ratio_plot}), where only statistical errors of the ratios are 
given. 
        
   The data show a strong 
decrease of the $\Lambda/\Sigma^0$ - production ratio in the excess energy 
range from $\sim 10~\mbox{MeV}$ to $\sim \mbox{20~MeV}$.
Above 20 MeV the slope is much smaller and the ratio seems to approach 
slowly the high energy level.

\begin{table}[h]
\caption{Cross section ratios for the $\Lambda$/$\Sigma^0$ productions at six 
excess energies.}
\label{cross_ratio}
 \begin{center}
 \begin{tabular}{|c|c|}
 \hline
Excess  energy                               &  \raisebox{-0.2cm}[0.35cm][0.3cm]{$\frac{\sigma(\pppKL)}{\sigma(\pppKSo)}$} \\
\raisebox{0.15cm}[0.2cm][0.1cm]{$ Q $ [MeV]} &\\
   \hline
13.9$\pm$0.6& 18.1 $\pm$ 4.1 \\
15.9$\pm$0.8& 15.5 $\pm$ 2.4 \\
20.3$\pm$0.8& 13.0 $\pm$ 2.6 \\
30.0$\pm$0.7& 10.9 $\pm$ 3.4 \\
39.7$\pm$1.2& 10.8 $\pm$ 2.3 \\
59.2$\pm$0.7& 8.0 $\pm$ 2.7 \\
  \hline
\end{tabular}
\end{center}
\vspace{-0.4cm}
\end{table}

\section{Comparison with model expectations\label{comparison}}

Within different meson exchange models attempts have been made to reproduce 
the behavior of the $\Lambda / \Sigma$ cross section ratio, the high
threshold value as well as the energy dependence.

A strong $\Sigma N \rightarrow \Lambda p$ 
final state conversion \cite{sew99} is
rather excluded as a dominant origin 
of the observed $\Sigma^0$ suppression --- at least according to a coupled 
channel calculation of the J\"ulich group~\cite{gas00}, where both $\pi$ and 
$K$ exchange are taken into account with inclusion of the final state 
interaction (FSI) effects. Here
$\Lambda$ production is dominated~\cite{gas00} by kaon exchange, which is  in 
line with the experimental results obtained by the DISTO 
collaboration~\cite{bal99} at higher excess energies 
($\mbox{Q} = 430\,\mbox{MeV}$), where the importance of $K$ exchange 
is confirmed by a measurement of the polarisation transfer coefficient.
Contrary, in case of $\Sigma^0$ production both $\pi$ and $K$ 
exchange are found to contribute with about the same strength~\cite{gas00}.
 
A destructive interference of the $\pi$ and $K$ exchange, suggested by 
Gasparian et al.~\cite{gas00,gas01}, is able to describe the large cross 
section ratio close--to--threshold.
Contributions from direct production as well as from heavy meson exchange were 
not considered in these calculations but might have an influence on the ratio 
of the $\Lambda/\Sigma^0$ production as suggested in 
Refs.~\cite{tsu97,tsu99,kai99a}.

\begin{figure}
\centerline{
\psfig{figure=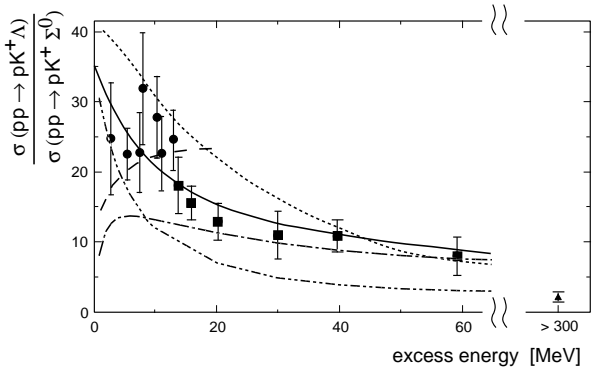,width=0.27\textwidth,angle=270}
}
\vspace{0.0cm}
\caption{Energy dependence of the cross section ratio for
$\Lambda/\Sigma^0$ production in proton--proton collisions. 
Circles and squares present the experimental data from~\cite{sew99}
and the present work, respectively.  The data point at 
$Q \ge 300~\mbox{MeV}$ is an average taken from the compilation of 
Ref.~\cite{bal88}.
The curves represent calculations within different models,
$\pi$ and $K$ exchange with destructive interference 
~\cite{gas01} (dashed line),
incoherent $\pi$ and $K$ exchange~\cite{sib00b} (dashed-dotted),
meson exchange with intermediate $N^*$ excitation ~\cite{sib00b}
(dashed-double-dotted) and
effective Lagrangian approach including $N^*$ excitation~\cite{shy01}
(dotted line). The solid line presents  the ratio of the fit functions 
given by formula~\ref{faldtwilkinflux} and shown in 
Fig.~\ref{cross_section_plot}.
}
\label{cross_ratio_plot}
\vspace{-0.6cm}
\end{figure}

In Fig.~\ref{cross_ratio_plot} the energy dependence of the cross
section ratio given by the different models is shown together with the
available data.
The predictions of the different models are of comparable quality even 
though different production mechanisms are considered. However, none of them
reproduces the data really well.

        Studies of the production ratio in Ref.~\cite{sib00b} consider two 
different models: The first one is based on $\pi$ and $K$ exchange 
calculations neglecting completely any interferences of the 
amplitudes~\cite{sib95} (dashed-dotted line). 
The second one considers the exchange of $\pi$ or 
heavier mesons with an excitation of intermediate $N^*$ resonances coupled to 
the $K^+ Y$ channel~\cite{tsu97,tsu99} (dashed-double-dotted line). 
Again, any interference of the 
amplitudes are neglected.

The resonance production is also taken into account in an effective Lagrangian 
approach~\cite{shy01} (dotted line) based on methods discussed in 
references~\cite{shy98,shy99,eng96},  where 
the stran\-ge\-ness production mechanism is modeled by the exchange of $\pi$, 
$\rho$, $\omega$ and $\sigma$ mesons with a subsequent excitation of the 
nucleon resonances $N^*(1650)$, $N^*(1710)$ and $N^*(1720)$. In those 
calculations the experimental data are reproduced within a factor of two.

The one-boson exchange calculation performed by La\-get~\cite{lag91,lag01} 
takes 
interference effects of pion and kaon exchanges  into account by selecting 
the relative sign for these two mechanism which gives the best description of
the cross sections. The results of those calculations not only reproduce 
the threshold data of the $\Lambda/\Sigma^0$ ratio within a factor of two and 
the polarisation transfer results of the DISTO experiment~\cite{bal99} but 
also describe the missing mass distribution obtained in 
the inclusive $K^+$ production measurements performed at SATURNE~\cite{sie94}.
However, since the predicted ratio is given only in the excess energy range
$4.5~\mbox{MeV} \le Q \le 16~\mbox{MeV}$~\cite{lag91,lag01}, 
the results are not shown in Fig.~\ref{cross_ratio_plot}.

The solid line in Fig.~\ref{cross_ratio_plot} results for the ratio of the fit
functions as shown in Fig.~\ref{cross_section_plot} which suggests that the 
energy dependence is simple governed by the different strengths of the $p$-$Y$ 
final state interactions but here the amplitudes are simply independently
normalized without considering the differences in the contributing production
amplitudes.

\section{Conclusion}

At the internal hydrogen cluster target facility $\mbox{COSY--11}$
the energy dependence of the total cross sections for 
the \pppKL and \pppKSo production 
was measured in the range of excess energies between 14 and 60 MeV
in order to investigate the transition of the 
$\Lambda/\Sigma^0$ cross section ratio  ${\cal R}$ from
${\cal{R}} \thickapprox 28$ at $Q \le 13~\mbox{MeV}$ to 
${\cal{R}}\thickapprox 2.5$ at $Q > 300~\mbox{MeV}$.

	A strong decrease of the cross section ratio in the excess energy 
range between 10 and 20 MeV is observed.
Various models are able to describe the data within a factor of two 
with comparable quality, even though they differ in the dominant contribution 
to the production mechanism.
%The data base in the hyperon sector is rather poor so that certain parameters
%like the $p-\Sigma^0$ interaction cannot be fixed giving some degree of 
%freedom in the calculations. 

Below $Q = 13~\mbox{MeV}$ the measured excitation functions were consistent 
with comparable final state interaction strengths in both channels, 
$p-\Lambda$ and $p-\Sigma^0$~\cite{sew99}.
The new data sugest  much weaker final state interactions in the 
$p-\Sigma^0$ channel than in the case of  $p-\Lambda$, at least  
as far as the present results are not feigned by contributions from higher 
partial waves and/or an energy dependence of the elementary production 
amplitude.

A measurement with high statistics at an excess energy of 
$40~\mbox{MeV} \leq Q \leq 60~\mbox{MeV}$
is highly desirable to study the angular distribution of the produced $\Lambda$
and $\Sigma^0$ hyperons. This would allow to determine whether higher partial 
waves already contribute in this energy range. 
Also, for a significant improvement of the accuracy of the information 
extracted from the $YN$ FSI a measurement of the corresponding invariant mass 
spectra would be very useful~\cite{gas03}.
%
%To improve the accuracy of the 
%extracted $p-\Sigma^0$ final state interaction strength a Dalitz plot analysis 
%is required which certainly calls for higher statistics data. This method was 
%proven to work in the investigations of the $p-\Lambda$ interaction close to 
%threshold~\cite{bal98b}.

In addition, data with polarized beam and hyperon polarization will be helpful
to disentangle the contributing exchange mesons. Also other isospin channels 
have to be considered. Calculations within the J\"ulich meson exchange 
model \cite{gas00} for other $\Sigma$ channels have shown that e.g. the 
reaction channel $pp \rightarrow n K^+\Sigma^+$ is strongly dependent 
on the $\pi-K$ interference. For a destructive interference the cross section 
for $pp \rightarrow n K^+\Sigma^+$ is expected to be a factor of three higher 
and for constructive interference a factor of three lower than the cross 
section for $pp \rightarrow p K^+\Sigma^0$.

Data of the reaction $pp \rightarrow n K^+\Sigma^+$  have already been 
taken~\cite{c11:prop} at COSY-11 and the analysis is in progress.

\section{Acknowledgements}

This work has been supported by the International B{\"uro} and the 
Verbundforschung of the BMBF, the Polish State Committee for Scientific 
Research,  the FFE grants from the Forschungs\-zentrum J{\"u}lich and the 
European Community -- Access to Research Infrastructure action of the 
Improving Human Potential Programme.
%
% BibTeX users please use
 \bibliographystyle{h-elsevier}
 \bibliography{bibliography}

\end{document}